# Algorithms for Wireless Capacity

Olga Goussevskaia, Magnús M. Halldórsson and Roger Wattenhofer

*Abstract*—In this paper we address two basic questions in wireless communication: First, how long does it take to schedule an arbitrary set of communication requests? Second, given a set of communication requests, how many of them can be scheduled concurrently? Our results are derived in an interference model with geometric path loss and consist of efficient algorithms that find a constant approximation for the second problem and a logarithmic approximation for the first problem. In addition, we analyze some important properties of the interference model and show that it is robust to various factors that can influence the signal attenuation. More specifically, we prove that as long as such influences on the signal attenuation are constant, they affect the capacity only by a constant factor.

## I. Introduction

Despite the omnipresence of wireless networks, surprisingly little is known about their algorithmic complexity and efficiency: Designing and tuning a wireless network is a matter of experience, regardless whether it is a WLAN in an office building, a GSM phone network, or a sensor network on a volcano.

We are interested in the fundamental communication limits of wireless networks. In particular, we would like to know what communication throughput can possibly be achieved. This question essentially boils down to spatial reuse, i.e., which devices can transmit concurrently, without interfering.

The answer to the question stated above depends, among other factors, on the topology of the network. One could be interested in networks where nodes are randomly distributed, or are positioned on a regular grid, as examples of best-case scenarios, i.e., where capacity is maximized. The problem of determining the capacity of such networks has been extensively studied, starting with the seminal work of Gupta and Kumar [21]. Another direction is to restrict attention to linksets with special properties. In [35] a power-assignment algorithm which schedules a strongly connected set of links in poly-logarithmic time was presented. This is probably the first algorithmic result in the physical model with guaranteed performance in worst-case topologies; it relies, however, crucially on the freedom to choose the links to be scheduled and that the connectivity requirement.

In this work we generalize this research to consider the capacity of *any* network: one with arbitrary topology and arbitrary set of communication requests. The computational aspect is fundamental: we need to be able to compute the capacity efficiently. Since general instances defy simple laws, the algorithm becomes the means to express capacity. Therefore, if one wants to know the capacity of any network, this paper provides the tool to do that, as it computes the capacity of any network up to a logarithmic factor.

An important issue when studying wireless networks is how to model interference. The most commonly used interference models can be roughly classified into graph-based models and fading channel models. Graph-based models, such as the protocol model, despite being a useful abstraction of wireless networks, are too simplistic. Consider for instance a case of three wireless transmissions, every two of which can be scheduled concurrently without a conflict. In a graph-based model one will conclude that all three transmissions may be scheduled concurrently as well, while in reality this might not be the case since wireless signals sum up. Instead, it may be that two transmissions together generate too much interference, hindering the third receiver from correctly receiving the signal of its sender. This many-to-many relationship makes understanding wireless transmissions difficult—a model where interference sums up seems paramount to truly comprehending wireless communication. Similarly, a graph-based model oversimplifies wireless attenuation. In graph-based models the signal is "binary", as if there was an invisible wall at which the signal immediately drops. Not surprisingly, in reality the signal decreases gracefully with distance.

Fading channel models, such as the *physical model* (formally introduced in Section III), offer a more realistic representation of wireless communication. A signal is received successfully if the SINR—the ratio of the received signal strength to the sum of the interference caused by all other nodes sending simultaneously, plus noise—is above a hardware-defined threshold. This definition of a successful transmission, as opposed to the graph-based definition, accounts also for interference generated by transmitters located far away. Observe that, since the SINR depends on combinations of the transmissions scheduled concurrently, interference is no longer a binary relation (or a graph). This makes the analysis of algorithms more challenging than in graph-based models.

The capacity of wireless networks in fading channel models has received a lot of attention from researchers in information, communication and network theory. In contrast to the results in graph-based models, which are of algorithmic nature and concerned with arbitrary instances, the results in the physical model have been typically based on heuristics evaluated by simulation of average scenarios. Analytical work in this context has been done only for special cases, e.g. when the network has a grid structure or when traffic is random. Therefore, these results give little insight into the computational complexity of the problem and cannot be translated into algorithms that can ultimately lead to new protocols.

In this paper we focus on a specific part of the problem

This work is based on the preliminary conference papers [16] and [25].

O. Goussevskaia is with the Department of Computer Science, Federal University of Minas Gerais, Brazil. Email: olga@dcc.ufmg.br

M. M. Halldórsson is with the School of Computer Science, Reykjavik University, Iceland. Email: mmh@ru.is

R. Wattenhofer is with the Computer Engineering and Networks Laboratory, ETH Zurich, Switzerland. Email: wattenhofer@ethz.ch

of determining the throughput capacity of a wireless network. We study the problem of scheduling one-hop communication requests without power control, i.e., we do not consider routing nor power control problems. The specific questions we address are two classic issues in wireless communication: Given a set of arbitrary communication requests, (i) how many of them can be scheduled concurrently, and (ii) how long does it take to schedule all of them? We can solve the first problem asymptotically optimally. The solution of the first problem then directly leads to an understanding of the second problem. In particular, it gives an approximation which is optimal up to a factor that is logarithmic in the number of requests. Note that we compute the network's capacity up to a small insecurity, whereas the complete understanding is out of bound, since the problem is NP-hard [18].

These results contribute to the core of wireless communication, since the problems being solved form the basic building blocks for upper-layer network functions, such as multi-hop and multi-rate communication. Routing can be solved by attaching the routing machinery to the one-hop solution. Moreover, latest breakthroughs in the power control problem [29] have been built on top of results presented in this work.

Our third contribution is a proof of robustness of the physical model with geometric path loss. One may argue that in reality path loss will not follow a perfect geometric pattern. Instead, various factors can affect the transmission, e.g., antenna gain may be higher in some directions, obstacles may influence attenuation, and noise may be location dependent. We show that as long as influences are constant, results will only be affected by a constant. As such, the physical model is *robust*. This result holds in a variety of settings, including power controlled transmissions.

## II. RELATED AND CURRENT RESULTS

Most work in wireless scheduling in the physical (SINR) model is of heuristic nature, e.g. [5], [9]. Only after the work of Gupta and Kumar [21] analytical results became *en vogue*, but only restricted to networks with a well-behaving topology and traffic pattern. On the one hand this restriction keeps the math involved tractable, on the other hand, it allows for presenting the results in a concise form, i.e., "the throughput capacity of a wireless network with $X$ and $Y$ is $Z$", where $X$ and $Y$ are some parameters defining the network, and $Z$ is a function of the network size. This area of research has been exceptionally popular, with a multi-dimensional parameter space (e.g. node distribution, traffic pattern, transport layer, mobility). An intrinsic problem with this line of research is that, in practice, networks often do not resemble the models studied here, so one cannot learn much about the capacity of an arbitrary network. Moreover, it is difficult to deduce protocols, since the results are not algorithmic.

Mathematical programming techniques can be used to formulate the capacity problem and various extensions, typically in the form of convex programming (see, e.g., [39]). The NP-hardness of the problem [18] tells us, however, that one can only hope to solve small instances using such formulations. Additionally, they don't give algorithmic insights that can lead to light-weight solution mechanisms.

In contrast, there is a body of algorithmic work, but mostly on graph-based models. Studying wireless communication in graph-based models commonly implies studying some variants of independent set, matching, or coloring, e.g. [31]. Although these algorithms present extensive theoretical analysis, they are constrained to the limitations of a model that ultimately abstracts away the nature of wireless communication. The inefficiency of graph-based protocols in the SINR model is well documented and has been shown theoretically as well as experimentally [20], [32], [36].

Algorithmic work in the SINR model is fairly new; to the best of our knowledge it was started just a few years ago [35]. In this paper Moscibroda and Wattenhofer present an algorithm that successfully schedules a set of links (carefully chosen to strongly connect an arbitrary set of nodes) in polylogarithmic time, even in arbitrary worst-case networks. In contrast to our work the links themselves are *not* arbitrary, but have structure that simplifies the problem. In a follow-up paper, Moscibroda, Wattenhofer and Zollinger [37] first define the link scheduling problem, whose single-shot variant is the focus of this paper. These concepts have been extended and applied to topology control [15], [37], sensor networks [33], combined scheduling and routing [7], ultra-wideband [27], and analog network coding [19], just to name a few. Apart from these papers, algorithmic SINR results also started appearing here and there, such as in a game theoretic or distributed algorithms context, e.g., [3], [4], [6], [17], [28], [38].

Previous to our work, few papers appeared that tackle the problem of scheduling arbitrary wireless links. Goussevskaia et al. [18] showed that the problem is NP-complete, and Moscibroda et al. [34] evaluated popular heuristics. Both papers also present approximation algorithms, with approximation ratios that depend on network parameters and can become linear in the network size.

Since the original publication of our work [16], numerous results have appeared on different aspects of scheduling in the SINR model. The scheduling problem with linear power assignment was treated by Fanghänel et al. [13], including a nearly-constant approximation. The combined problem of scheduling and power control has been treated in [12], [23], [24], culminating in a recent constant factor approximation by Kesselheim [29]. Online algorithms for the dynamic scheduling problem, where communication requests arrive dispersed over time, have been examined in [10], [11], [22]. Game theory was treated in [1], [2], [8], and auctioning of spectrum in [26]. Distributed algorithms have been proposed in [2], [8], [30].

Recently, Halldórsson and Mitra [24] have extended the results of this paper in two ways: from the Euclidean plane to general metrics, and to a more general range of fixed power assignments.

### A. Our Results

In this paper we present the first results that provide approximation guarantees independent of the topology of the network. Our main contributions are:

- Given an arbitrary set of requests, we present a simple greedy algorithm that chooses a subset of the requests

that can be transmitted concurrently without violating the SINR constraints. This subset is guaranteed to be within a constant factor of the optimal subset.

- Furthermore, by applying the single-slot subroutine repeatedly we realize a $O(\log n)$-approximation for the problem of minimizing the number of time slots needed to schedule a given set of arbitrary requests. Simulation results indicate that this approximation algorithm, besides having an exponentially better approximation ratio in theory, is also practical. It is easy to implement and achieves superior performance in various network scenarios.
- We also present a non-approximability result for the scheduling problem in the non-geometric SINR model. More specifically, we show that in the SINR model where path-loss is set arbitrarily (i.e., not determined by the Euclidean coordinates of the nodes), it is NP-hard to approximate the scheduling problem to within $n^{1-\varepsilon}$ factor, for any constant $\varepsilon > 0$.
- Finally, we present a general robustness result, showing that constant parameter and model changes will modify the result only by a constant.
- All our results rely on a new definition to understand physical interference: *affectance*. This definition has been proved to be of general utility for analyzing algorithms in the SINR context, both for scheduling with fixed-but-different power assignments [30], [24] and in power controlled scheduling [23], [29], [24].

One may argue that media access and scheduling are fundamental problems when it comes to wireless communication. Although power controlled cases are interesting from a theoretical point of view, practically the most important cases are those with constant power. Although there are many actual wireless networks where nodes can choose different transmission powers, the selection is then either restricted to a small set of possible power levels, or a bounded power range. The analytical results of this paper hold for both extensions. Apart from constants, all our findings are directly transferrable to bounded power set and to bounded ratio of maximum and minimum power. As such we believe that our results are practically relevant.

The main features of the current paper, including the general style of the algorithm, affectance analysis, and signal strengthening, factor in and influence nearly all recent work.

This paper fixes several minor plus one larger mistake (an erroneous claim on the scheduling complexity in [25]) from the preliminary conference versions [16] and [25].

## III. NOTATION AND MODEL

Given is a set of links $L = \{\ell_1, \ell_2, \ldots, \ell_n\}$, where each link $\ell_v$ represents a communication request from a sender $s_v$ to a receiver $r_v$. We assume the senders and receivers are points in the Euclidean plane; this can be extended to other metrics. The Euclidean distance between two points $p$ and $q$ is denoted $d(p,q)$. The asymmetric distance from link $v$ to link $w$ is the distance from $v$'s sender to $w$'s receiver, denoted $d_{vw} = d(s_v, r_w)$. The length of link $\ell_v$ is denoted $d_{vv} = d(s_v, r_v)$. We shall assume for simplicity of exposition that all links are of different length; this does not affect the results. We assume that each link has a unit-traffic demand, and model the case of non-unit traffic demands by replicating the links. We also assume that all nodes transmit with the same power level $P$. We show later how to extend the results to variable power levels, with a slight increase in the performance ratio.

We assume the *path loss radio propagation* model for the reception of signals, where the received signal from transmitter $w$ at receiver $v$ is $P_{wv} = P/d_{wv}^\alpha$ and $\alpha > 2$ denotes the path-loss exponent. We adopt the *physical interference model*, in which a node $r_v$ successfully receives a message from a sender $s_v$ if and only if the following condition holds:

$$\frac{P_{vv}}{\sum_{\ell_w \in S \setminus \{\ell_v\}} P_{wv} + N} \geq \beta \;, \qquad (1)$$

where $N$ is the ambient noise, $\beta$ denotes the minimum SINR (signal-to-interference-plus-noise-ratio) required for a message to be successfully received, and $S$ is the set of concurrently scheduled links in the same channel or *slot* as $\ell_v$. We say that $S$ is *SINR-feasible* if (1) is satisfied for each link in $S$.

The problems we treat are the following. In all cases are we given a set of links of arbitrary lengths. In the Scheduling problem, we want to partition the set of input links into minimum number of SINR-feasible sets, each referred to as a *slot*. In the Single-Shot Scheduling problem, we seek the maximum cardinality subset of links that is SINR-feasible.

We make crucial use of the following new definitions.

*Definition 3.1:* The *relative interference* (RI) of link $\ell_w$ on link $\ell_v$ is the increase caused by $\ell_w$ in the inverse of the SINR at $\ell_v$, namely $RI_w(v) = P_{wv}/P_{vv}$. For convenience, define $RI_v(v) = 0$. Let $c_v = \frac{1}{1 - \beta N/P_{vv}}$ be a node-dependent constant that indicates the extent to which the ambient noise approaches the required signal at receiver $r_v$. The *affectance* of link $\ell_v$, caused by a set $S$ of links, is the sum of the relative interferences of the links in $S$ on $\ell_v$, scaled by $c_v$, or

$$a_S(\ell_v) = c_v \cdot \sum_{\ell_w \in S} RI_w(v) \;.$$

For a single link $\ell_w$, we use the shorthand $a_w(\ell_v) = a_{\{\ell_w\}}(\ell_v)$. We define a *p-signal* set or schedule to be one where the affectance of any link is at most $1/p$.

The constant $c_v$ is monotone increasing with the length of the link: $d_{vv} \geq d_{ww}$ implies that $c_v \geq c_w$. Note that $c_v \geq 1$, with equality holding only in the absence of noise.

*Observation 3.2:* The affectance function satisfies the following properties for a set $S$ of links:

1) *(Range)* $S$ is SINR-feasible if and only if, for all $\ell_v \in S$, $a_S(\ell_v) \leq 1/\beta$.
2) *(Additivity)* $a_S = a_{S_1} + a_{S_2}$, whenever $(S_1, S_2)$ is a partition of $S$.
3) *(Distance bound)* $a_w(\ell_v) = c_v \cdot \left(\frac{d_{vv}}{d_{wv}}\right)^\alpha$, for any pair $\ell_w, \ell_v$ in $S$.

Note that the concepts of affectance and relative interference are equally useful in contexts of non-uniform power assignments. If $P_v$ is the power of link $\ell_v$, the affectance of link $\ell_v$ on $\ell_v$ is given by $a_w(\ell_v) = c_v \cdot \left(\frac{P_w/d_{wv}}{P_v/d_{vv}}\right)$.

## IV. Properties of SINR Schedules

We present in Section IV-A properties of schedules in the SINR model, which double as tools for the algorithm designer. Then in Section IV-B, we examine the desirable property of link dispersion, and how any schedule can be dispersed at a limited cost.

### A. Robustness of the SINR Model

We now explore how signal requirements (in the value of $\beta$), or equivalently interference tolerance, affects schedule length. It is not *a priori* obvious that minor discrepancies cause only minor changes in schedule length, but by showing that it is so, we can give our algorithms the advantage of being compared with a stricter optimal schedule. This also has implications regarding the robustness of SINR models with respect to perturbations in signal transmissions.

The pure geometric version of SINR given in (1) is an idealization of true physical characteristics. It assumes, e.g., perfectly isotropic radios, no obstructions, and a constant ambient noise level. That begs the question, why move algorithm analysis from analytically amenable graph-based models to a more realistic model if the latter isn't all that realistic? Fortunately, the fact that schedule lengths are relatively invariant to signal requirements shows that these concerns are largely unnecessary.

The results of this section apply equally to scheduling links of different powers. It also applies to throughput optimization.

*Theorem 4.1:* There is a polynomial-time algorithm that takes a $p$-signal schedule and refines it into a $p'$-signal schedule, for $p' > p$, increasing the number of slots by a factor of at most $\lceil 2p'/p \rceil^2$.

  *Proof:* Consider a $p$-signal schedule $\mathcal{S}$ and a slot $S$ in $\mathcal{S}$. We partition $S$ into a sequence $S_1, S_2, \ldots$ of sets. Order the links in $S$ in decreasing order. For each link $\ell_v$, assign $\ell_v$ to the first set $S_j$ for which $a_{S_j}(\ell_v) \leq 1/(2p')$, i.e. the accumulated affectance on $\ell_v$ among the previous, longer links in $S_j$ is at most $1/(2p')$. Since each link $\ell_v$ originally had affectance at most $1/p$, then by the additivity of affectance, the number of sets used is at most $\lceil \frac{1/p}{1/(2p')} \rceil = \lceil \frac{2p'}{p} \rceil$.

We then repeat the same approach on each of the sets $S_i$, processing the links this time in increasing order. The number of sets is again $\lceil \frac{2p'}{p} \rceil$ for each $S_i$, or $\lceil \frac{2p'}{p} \rceil^2$ in total. In each final slot (set), the affectance on a link by shorter links in the same slot is at most $1/2p'$. In total, then, the affectance on each link is at most $2 \cdot 1/2p' = 1/p'$. ∎

This result applies in particular to optimal solutions. Let $\psi(L)$ denote the minimum number of slots in an SINR-feasible schedule of a linkset $L$, and let $\psi_p(L)$ denote the same quantity for an optimal $p$-signal schedule. It is not *a priori* clear that a smooth relationship exists between $\psi_p$ and $\psi = \psi_\beta$, for $p > 1$.

*Corollary 4.2:* For any linkset $L$ and any $p > 1$, $\psi_p(L) \leq \lceil 2p/\beta \rceil^2 \psi(L)$.

This has significant implications. One regards the validity of studying the pure SINR model. As asked in [16], "what if the signal is attenuated by a certain factor in one direction but by another factor in another direction?" A generalized physical model was introduced in [37] to allow for such a deviation. Theorem 4.1 implies that scheduling is relatively robust under discrepancies in the SINR model. This validates analytic studies of the pure SINR model, in spite of its simplifying assumptions.

*Corollary 4.3:* If a scheduling algorithm gives a $\rho$-approximation in the SINR model, it provides a $O(\theta^2 \rho)$-approximation in variations in the SINR model with a discrepancy of up to a factor of $\theta$ in signal attenuation or ambient noise levels.

This result can be contrasted with the result of Section VII, that shows a strong $n^{1-\epsilon}$-approximation hardness of scheduling in an abstract (non-geometric) SINR model that allows for arbitrary distances between nodes. Alternatively, Theorem 4.1 allows us to analyze algorithms under more relaxed situations than the optimal solutions that we compare to.

It is important to note that these results do not depend on the power assignment and apply equally well in the power-control setting. Also, they actually do not depend on the formula used to compute affectance or relative interference, and apply also in non-geometric and non-metric settings.

**Remark**: Note that the converse of Theorem 4.1 – that a schedule can be shortened by a constant factor so that the signal decreases only by a constant factor – does not hold. An easy example is found by duplicating a feasible set $S$ by any number $t$ of copies (possibly separating the nodes by a sufficiently small distance). Any attempt to use fewer than $t$ slots results in an arbitrarily bad signal.

### B. Dispersion properties

One desirable property of schedules is that links in the same slot be spatially well separated. This blurs the difference in position between sender and receiver of a link, since it affects distances only by a small constant. Intuitively, we want to measure nearness as a fraction of the lengths of the respective links. Given the affectance measure, it proves to be useful to define nearness somewhat less restrictively.

*Definition 4.4:* Link $\ell_w$ is said to be $q$-near link $\ell_v$, if $d_{wv} < q \cdot c_v^{1/\alpha} \cdot d_{vv}$. A set of links is $q$-dispersed if no (ordered) pairs of links in the set are $q$-near.

Observation 3.2, item 3, states that link $\ell_w$ is $q$-near a link $\ell_v$ if and only if $a_w(\ell_v) > q^{-\alpha}$. This immediately gives the following strengthening of Lemma 4.2 in [16].

*Lemma 4.5:* Fewer than $q^\alpha / \beta$ senders in an SINR-feasible set $S$ are $q$-near to any given link $\ell_v \in S$.

For constant $q, \alpha$, any schedule can be made dispersed at a cost of a constant factor.

*Lemma 4.6:* There is a polynomial-time algorithm that takes a SINR-feasible schedule and refines it into a $q$-dispersed schedule, increasing the number of slots by a factor of at most $\lceil (q+2)^\alpha \rceil$.

  *Proof:* Let $S$ be a slot in the schedule. We show how to partition $S$ into sets $S_1, S_2, \ldots, S_t$ that are $q$-dispersed, where $t \leq (q+2)^\alpha + 1$.

Process the links of $S$ in increasing order of length, assigning each link $\ell_v$ "first-fit" to the first set $S_j$ in which the receiver $r_v$ is at least $\left(q c_v^{1/\alpha} + 2\right) \cdot d_{vv}$ away from any other link. Let $\ell_w$ be a link previously in $S_j$, and note that $\ell_w$ is shorter than $\ell_v$. By the selection rule, $d_{wv} \geq \left(qc_v^{1/\alpha}+2\right) \cdot d_{vv} > qc_v^{1/\alpha} \cdot d_{vv}$. Also,

$$\begin{aligned} d_{vw} &\geq d_{wv} - d_{ww} - d_{vv} \\ &\geq \left(qc_v^{1/\alpha}+1\right)d_{vv} - d_{ww} \\ &\geq qc_v^{1/\alpha}d_{ww} \\ &\geq qc_w^{1/\alpha}d_{ww}. \end{aligned}$$

Since this holds for every pair in the same set, the schedule is $q$-dispersed. Suppose $S_t$ is the last set used by the algorithm, and let $\ell_v$ be a link in it. Then, each $S_i$, for $i=1,2,\ldots,t-1$, contains a link whose sender is closer than $(qc_v^{1/\alpha}+2)\cdot d_{vv} \leq (q+2)c_v^{1/\alpha}d_{vv}$ to $r_v$, i.e., is $(q+2)$-near to $\ell_v$. By Lemma 4.5, $t-1 < (q+2)^\alpha/\beta$. Hence, $t \leq \lceil (q+2)^\alpha/\beta \rceil$. ∎

Intuitively, there is a correlation between low affectance and high dispersion in schedules. The following result makes this connection clearer. The converse is, however, not true, since high interference can be caused by shorter far-away links.

*Lemma 4.7:* A $p$-signal schedule is also $p^{1/\alpha}$-dispersed.

*Proof:* Let $\ell_v$ and $\ell_w$ be an ordered pair of links in a slot $S$ in a $p$-signal schedule. By definition, $a_w(\ell_v) \leq a_S(\ell_v) \leq 1/p$. By Observation 3.2, item 3, $d_{wv} \geq (c_v p)^{1/\alpha} \cdot d_{vv}$. ∎

We remark that the results given in this subsection apply only to uniform power assignments, unlike the previous subsection.

## V. Approximation Algorithms

We now give a constant-factor approximation algorithm for Single-Shot Scheduling. We aim for conceptual simplicity, rather than optimizing the constants.

Let $C = 2^3 9 = 72$, $\tau = 2 + \max\left(2, \left((C+1)\beta\frac{\alpha-1}{\alpha-2}\right)^{\frac{1}{\alpha}}\right)$, and $c = 1/\tau^\alpha$.

**A**($c$)
  sort the links $\ell_1, \ell_2, \ldots, \ell_n$ by non-decreasing order of length
  $S \leftarrow \emptyset$
  for $v \leftarrow 1$ to $n$ do
    if ($a_S(\ell_v) \leq c$)
      add $\ell_v$ to $S$
  output $S$

It is rather surprising that a $O(1)$-approximation algorithm can be obtained in a single sweep. This should help in applying the ideas further, e.g., in distributed implementations. Note that recent research shows that such a single sweep is also feasible when using power control [29].

It is not immediate that algorithm **A** produces a feasible solution.

*Lemma 5.1:* Algorithm **A** produces a $(\tau - 2)$-dispersed solution.

*Proof:* Let $\ell_w$ be a link in the set $S$ output by algorithm **A**. Let $N_w^-$ ($N_w^+$) be the set of links in $S$ that are shorter (longer) than $\ell_w$. Consider first a link $\ell_u \in N_w^-$. Since $\ell_w$ was added by the algorithm after adding $\ell_u$, $a_u(\ell_w) \leq c = 1/\tau^\alpha$, which implies by Observation 3.2, item 3, that $d_{uw} \geq \tau c_w^{1/\alpha}d_{ww} > (\tau-2)c_w^{1/\alpha}d_{ww}$. Consider next a link $\ell_v \in N_w^+$. Since $\ell_v$ was added after $\ell_w$, it holds that $a_w(\ell_v) \leq c = 1/\tau^\alpha$. So, by Observation 3.2, $d_{wv} \geq \tau \cdot c_v^{1/\alpha}d_{vv}$. Recall that $c_v \geq c_w$ whenever $d_{vv} \geq d_{ww}$. Then, using the triangular inequality,

$$\begin{aligned} d_{vw} &= d(s_v, r_w) \geq d_{wv} - d_{vv} - d_{ww} \\ &\geq \left(\tau c_v^{1/\alpha} - 2\right)d_{vv} \\ &\geq (\tau-2)c_w^{1/\alpha}d_{ww}. \end{aligned}$$

Since this holds for every ordered pair in $S$, we have that $S$ is $(\tau-2)$-dispersed. ∎

*Lemma 5.2:* Let $S$ be a $Z$-dispersed feasible set of links, where $Z \geq 2$. Then, for any link $\ell_v$ in $S$, it holds that

$$a_{S_v^+}(\ell_v) < \left(\frac{\alpha-1}{\alpha-2}C\right)Z^{-\alpha},$$

where $S_v^+$ is the set of links in $S$ at least as long $\ell_v$.

*Proof:* Let $z = Zc_v^{1/\alpha}$. Form a disc $D_w$ of radius $r = (z-1)d_{vv}/2$ around each sender $s_w$ in $S_v^+$. We claim that these discs are disjoint. By the dispersion property, the distance from any sender $s_u \in S$ to any receiver $r_w \in S_v^+$, $w \neq u$, is at least $Zc_w^{1/\alpha}d_{ww} \geq zd_{ww}$, using that $c_w \geq c_v$ since $\ell_w \geq \ell_v$. It follows by the triangular inequality that the separation between two senders $s_u, s_w$ in $S$ is at least $(z-1)d_{ww} \geq (z-1)d_{vv} = 2r$, and thus the discs are disjoint.

We next partition the sender set in $S_v^+$ into concentric rings $R_k$ of width $z \cdot d_{vv}$ around the receiver $r_v$. Each ring $R_k$ contains all senders $s_w \in S_v^+$ satisfying $k(z \cdot d_{vv}) \leq d_{wv} \leq (k+1)(z \cdot d_{vv})$. We know that the first ring $R_0$ contains no sender (since such links would be incompatible with $\ell_v$). For each $k > 0$, the senders in $R_k$ are contained in an annulus $A_k$ centered at $r_v$ of width $zd_{vv} + 2r = (2z-1)d_{vv}$ that has $r$ added both to the inside and outside of $R_k$. The area of $A_k$ is

$$\begin{aligned} \text{Area}(A_k) &= \left[(d_{vv}(k+1)z+r)^2 - (d_{vv}kz-r)\right]\pi \\ &= (2k+1)d_{vv}^2 z(2z-1)\pi. \end{aligned}$$

Since discs $D_w$ of area $\text{Area}(D_w) = r^2\pi$ around senders in $S_v^+$ do not intersect, and the minimum distance between $r_v$ and $s_w \in R_k, k > 0$ is $k(z \cdot d_{vv})$, we can use an area argument to bound the number of senders inside each ring. The total relative interference from senders in $R_k, k \geq 1$ on $\ell_v$ is bounded by

$$\begin{aligned} RI_{R_k}(\ell_v) &\leq \sum_{s_w \in R_k} RI_{s_w}(\ell_v) \\ &\leq \frac{A(A_k)}{A(D_w)} \cdot \frac{1}{(kz)^\alpha} \\ &\leq \frac{(2k+1)}{k^\alpha} \cdot \frac{4}{z^\alpha}\frac{z(2z-1)}{(z-1)^2} \\ &\leq \frac{1}{k^{(\alpha-1)}} \cdot \frac{2^3 9}{z^\alpha}, \end{aligned}$$

where the last inequality holds since $k \geq 1 \Rightarrow 2k+1 \leq 3k$ and $z \geq 2 \Rightarrow z-1 \geq z/2$ and $2z-1 \leq 3(z-1)$. Summing up the interferences over all rings yields

$$RI_{S_v^+}(\ell_v) < \sum_{k=1}^\infty RI_{R_k}(\ell_v) \leq \sum_{k=1}^\infty \frac{1}{k^{\alpha-1}} \cdot \frac{C}{z^\alpha} < \frac{\alpha-1}{\alpha-2} \cdot \frac{C}{z^\alpha},$$

where the last inequality holds since $\alpha > 2$. This results in affectance of

$$a_{S_v^+}(\ell_v) = c_v RI_{S_v^+}(\ell_v) < \frac{\alpha - 1}{\alpha - 2} C \cdot \left(\frac{c_v^{1/\alpha}}{z}\right)^\alpha ,$$

as claimed. ∎

*Theorem 5.3:* Algorithm **A** produces an SINR-feasible solution.

*Proof:* Let $\ell_w$ be a link in the set $S$ output by algorithm **A**. Let $S_w^-$ ($S_w^+$) be the set of links in $S$ that are shorter (longer) than $\ell_w$. The links in $S_w^-$ were processed before $\ell_w$, so by the if-condition in the algorithm, $a_{S_w^-}(\ell_v) \leq c$. Note that $c \leq \frac{1}{(C+1)\beta}$. By Lemma 5.1, $S$ is $\tau - 2$-dispersed, so by Lemma 5.2 and the definitions of $\tau$ and dispersion,

$$a_{S_w^+}(\ell_w) < \left(\frac{\alpha - 1}{\alpha - 2} C\right) \frac{1}{(\tau - 2)^\alpha} \leq \frac{C}{(C+1)\beta}.$$

Hence, the affectance of each link in $S$ is at most $a_{S_w^-}(\ell_v) + a_{S_w^+}(\ell_v) \leq 1/\beta$. ∎

### A. Performance analysis

*Definition 5.4:* Let $\mathcal{R}$ and $\mathcal{B}$ be disjoint pointsets in a metric space $(\mathcal{V}, d)$, referred to as the *red* and *blue* points, respectively. A point $b \in \mathcal{B}$ is *blue-dominant* if every ball $B_\delta(b) = \{w \in \mathcal{B} | d(w, b) \leq \delta\}$ around $b$ contains more blue points than red points. Formally, $\forall \delta \in \mathbb{R}_0^+ : |B_\delta(b) \cap \mathcal{B}| > |B_\delta(b) \cap \mathcal{R}|$.

For a red point $r \in \mathcal{R}$ and a set $G \subseteq \mathcal{B}$ of blue points, we say that $G$ *guards* $r$ if for all $b \in \mathcal{B} \setminus G$, we have that $B_{d(b,r)}(b) \cap G \neq \emptyset$.

*Lemma 5.5:* (*Blue-dominant centers lemma*) Let $\mathcal{R}$ and $\mathcal{B}$ be disjoint sets of red and blue points in a 2-dimensional Euclidean space. If $|\mathcal{B}| > 5 \cdot |\mathcal{R}|$ then there exists at least one blue-dominant point in $\mathcal{B}$.

*Proof:* Process the points in $\mathcal{R}$ in an arbitrary order while maintaining a subset $\mathcal{B}'$ of $\mathcal{B}$ as follows (initially, $\mathcal{B}' = \mathcal{B}$). For each $r \in \mathcal{R}$, we construct a guarding set $G(r) \subseteq \mathcal{B}'$ (guarding $r$ relative to the current $\mathcal{B}'$) and remove $G(r)$ from $\mathcal{B}'$.

We claim that it is possible to construct a guarding set $G(r)$ of size at most 5. The procedure to construct $G(r)$ is as follows. Consider a red point $r$. Include a closest blue point $b_1 \in \mathcal{B}'$ in $G(r)$. Draw five sectors originating at $r$ in the following manner. The first sector has 120° and is centered at $b_1$, the remaining four sectors have 60° each and evenly divide the remaining area around $r$. For each of these four sectors $sec_j$, include the closest blue point $b_j \in sec_j$ in $G(r)$ (if $sec_j$ has no blue points from $\mathcal{B}'$, skip this sector). Now $G(r)$ has size at most 5 and we claim that it is guarding $r$. Suppose not. Then, there is a point $b^* \in \mathcal{B}' \setminus G(r)$ with $B_{d(b^*,r)}(r) \cap G(r) = \emptyset$. Suppose $b^*$ is located in $sec_j$ and we selected blue point $b_j$ from $sec_j$ into $G(r)$. This means that $d(b^*, b_j) > d(b^*, r)$, which implies that the sector angle is larger than 60°. (Note that if $G(r)$ contains no point $b_j$ from sector $sec_j$, then $b^*$ would have been picked to guard $r$ in that sector, also establishing a contradiction.)

After going through all the points in $\mathcal{R}$, the set $\mathcal{B}'$ is still nonempty by the assumption on the relative sizes of $\mathcal{R}$ and $\mathcal{B}$. We claim that every point in $\mathcal{B}'$ is now blue-dominant. This holds since (1) the guarding sets of point in $\mathcal{R}$ are pairwise disjoint and (2) every ball $B_\delta(b^*), b^* \in \mathcal{B}'$, that contains a red point $r$, contains also a blue point in $G(r)$. Hence, for every blue node $b^* \in \mathcal{B}'$, every ball $B_\delta(b^*)$ contains more blue points than red points ("more", since the center $b^*$ is also blue). ∎

*Lemma 5.6:* Let $\nu = 2(3\tau/2)^\alpha$ be a constant. Let $ALG$ be the solution output by algorithm **A** on the given instance and $OPT_\nu$ be an optimal $\nu$-signal solution. Then, $|OPT_\nu| \leq 5|ALG|$.

*Proof:* Let $\mathcal{R} = \{s_w | \ell_w \in ALG \setminus OPT_\nu\}$ and $\mathcal{B} = \{s_v | \ell_v \in OPT_\nu \setminus ALG\}$ be the sets of senders in exactly one of $ALG$ and $OPT_\nu$; we call them red and blue points, respectively. Suppose the claim is false. It follows that $|\mathcal{B}| > 5|\mathcal{R}|$. By Lemma 5.5, there is a blue-dominant $s_b$ in $\mathcal{B}$. We shall argue that the blue link $\ell_b = (s_b, r_b)$ would have been picked by our algorithm, which is a contradiction.

Consider any red point $s_x \in \mathcal{R}$. Let $D = d(s_x, s_b)$. Let $s_y$ denote the guard for $s_x$ w.r.t. $s_b$, i.e., the blue point that is closer to $s_b$ than $s_x$ is, i.e., within distance $D$ from $s_b$. Note that by Lemma 4.7, $OPT_\nu$ is a $s$-dispersed set, where $s = \nu^{1/\alpha} \geq 3\tau/2 \geq 6$. Applying Definition 4.4, we know that $d_{xb} \geq s \cdot c_v^{1/\alpha} \cdot d_{bb}$. Using $c_v \geq 1$, we get $d_{xb} \geq 6 d_{bb}$. The guarding property and the triangular inequality ensure that

$$d_{yb} \leq d(s_y, s_b) + d_{bb} \leq D + d_{bb} \leq d_{xb} + 2 d_{bb} \leq \frac{4}{3} d_{xb} .$$

Thus,

$$a_x(b) = c_b \left(\frac{d_{bb}}{d_{xb}}\right)^\alpha \leq c_b \left(\frac{4}{3} \cdot \frac{d_{bb}}{d_{yb}}\right)^\alpha = \left(\frac{4}{3}\right)^\alpha a_y(b) .$$

Let $t$ denote $\left(\frac{3}{4}\right)^\alpha$. This holds for any $s_x \in \mathcal{R}$, so the total interference that $\ell_b$ receives from the red senders (those in $ALG$) is at least $t$ times that from the blue senders. Since $\ell_b$ is in $OPT_\nu$, it is affected by at most $1/\nu$ by $OPT_\nu$. Using that each node in $OPT_\nu$ participates in at most one guardset, we get that

$$\begin{aligned} a_{ALG \setminus OPT_\nu}(\ell_b) &= \sum_{s_x \in \mathcal{R}} a_x(\ell_b) \\ &\leq \sum_{\ell_g \in \mathcal{B}} t \cdot a_g(b) \\ &= t \cdot a_{OPT \setminus ALG}(\ell_b) \\ &\leq t/\nu < c/2. \end{aligned}$$

Further, since $OPT_\nu$ is a $\nu$-signal solution, $a_{ALG \cap OPT_\nu}(\ell_b) \leq 1/\nu < c/2$. Thus,

$$a_{ALG}(\ell_b) = a_{ALG \setminus OPT_\nu}(\ell_b) + a_{ALG \cap OPT_\nu}(\ell_b) < c ,$$

which contradicts the assumption that $\ell_b$ was not selected by the algorithm. ∎

The following result is now immediate from Lemma 5.6 in combination with the correctness result in Theorem 5.3 and the signal-strengthening property of Cor. 4.2.

*Theorem 5.7:* Algorithm **A** approximates the Single-Shot Scheduling problem within a constant factor.

## B. Scheduling approximation

Given the constant factor approximation for the Single-Slot Scheduling problem, we get a $O(\log n)$-approximation for the Scheduling problem by repeatedly executing the Single-Slot Scheduling algorithm, and as such always removing a large set of links that can be scheduled concurrently, without interference.

*Theorem 5.8:* Repeated application of algorithm **A** yields an $O(\log n)$-approximation for the Scheduling problem.

*Proof:* Recall that $\psi$ is the minimum number of slots in a feasible solution, and let $\rho = O(1)$ be the performance guarantee of **A**. Any subset $S'$ of the input instance with $N$ links contains a feasible set of size $N/\psi$. Thus, Algorithm **A** applied to $S'$ results in a feasible subset of size at least $N/(\rho\psi)$, with the number of remaining unscheduled links becoming at most $N(1 - 1/(\rho\psi))$. Starting with $n$ links, the number of unscheduled links remaining after $s$ iterations is at most $n(1-1/(\rho\psi))^s < ne^{-s/(\rho\psi)}$. Thus, when $s \geq \ln n \cdot \rho\psi$, less than one link remains unscheduled, that is, all the links have been scheduled. Hence, $\ln n \cdot \rho\psi$ slots suffice, for an approximation factor of $\rho \ln n$. ∎

*a) Handling different transmission powers:* We can treat the case when links transmit with different powers in two different ways. Let $P_{max}$ ($P_{min}$) be the maximum (minimum) power used by a link, respectively. By introducing a factor of $P_{min}/P_{max}$ into the affectance threshold $c$, our algorithm still produces a feasible schedule, that is longer by a factor of at most $P_{max}/P_{min}$.

Alternatively, we can partition the instance into "power regimes", where each regime consists of links whose powers are equal up to a factor of 2. We schedule each power regime separately, obtaining an approximation factor of at most $\log P_{max}/P_{min}$, or at most the number of different power values.

If $P_{max}/P_{min}$ cannot be bounded, and if more generally the number of power levels cannot be bounded, we refer to recent work of [24] and [29].

## VI. SIMULATION RESULTS

In this section we present some simulation results (first given in [16]) to better illustrate the practical appeal of the scheduling approximation algorithm (to which we refer as ApproxLogN). We evaluate a heuristic improvement of Algorithm A, where $\hat{c} = \max(2, (2^5 3^2 \beta(\alpha-1)/(\alpha-2))^{1/\alpha})$.

**B**($c$)
    sort the links $\ell_1, \ell_2, \ldots, \ell_n$ by non-decreasing order of length
        $S \leftarrow \emptyset$
        for $v \leftarrow 1$ to $n$ do
          if ($a_S(\ell_v) \leq 2/3$ and $d_{wv} > \hat{c} \cdot d_{ww}, \forall \ell_w \in S$)
            add $\ell_v$ to $S$
    output $S$

We generated two kinds of topologies: *random* and *clustered* (see Figures 1(a) and 1(b)). In the random topology, $n$ receiver nodes were distributed uniformly at random on a plane field of size 1000x1000 units, and $n$ senders were positioned uniformly at random inside discs of radius $l_{max}$ around each of the receivers. In the clustered topology, $n_C$ cluster center positions were selected uniformly at random on the plane, and $n/n_C$ sender-receiver pairs were positioned uniformly at random inside discs of radius $r_C$ around each of them. The clustered topology aims to simulate a scenario of heterogeneous density distribution.

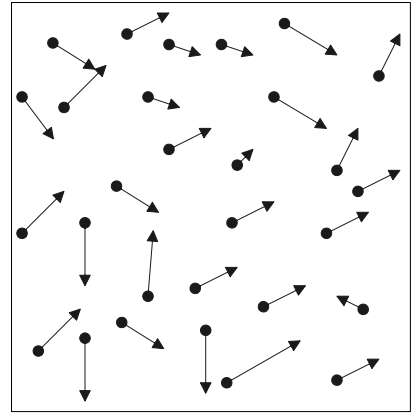

(a) Random.

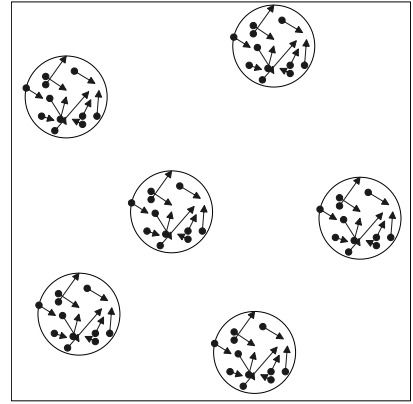

(b) Clustered.

Fig. 1. Simulated topologies: 1Kx1K field, $\alpha = 3$, $\beta = 1.2$, $N = 0$

We compare the performance of ApproxLogN to the performance of two other scheduling algorithms: GreedyPhysical (proposed in [5]) and ApproxDiversity (proposed in [18]). As ours, both are polynomial-time algorithms, specifically designed for the SINR model. In all experiments, the number of simulations was chosen large enough to obtain sufficiently small confidence intervals.

Firstly, we analyze the lengths of the schedules as a function of the number of nodes ($n \in \{100 \cdot 2^0, 100 \cdot 2^1, \cdots, 100 \cdot 2^8\}$). In Figures 2(a) and 2(b) the results for the random topology are shown. Since this scenario is not very challenging, all three algorithms have good performance, computing schedules of comparable sizes. GreedyPhysical presents slightly better performance in very low density scenarios (less than 1600 nodes). As the density increases, however, ApproxLogN presents increasingly better relative performance. In high densities (25.6K nodes) it computes, on average, 50% shorter schedules than GreedyPhysical and 2.5 times shorter schedules than ApproxDiversity.

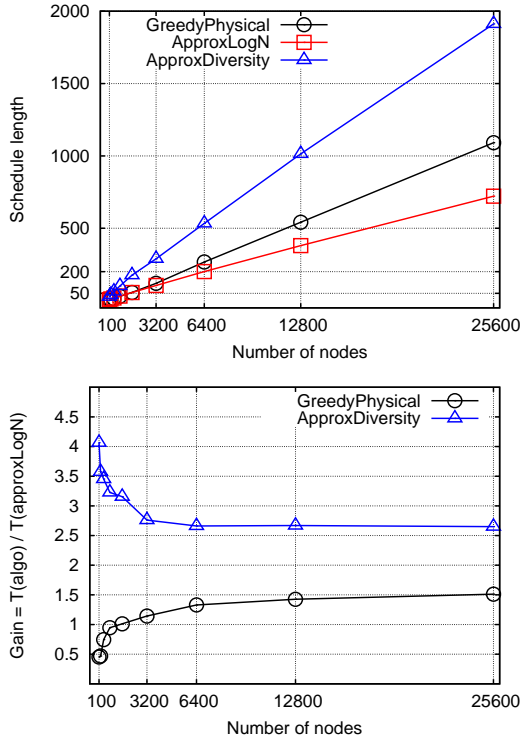

Fig. 2. Random Topology: $l_{max} = 20$.

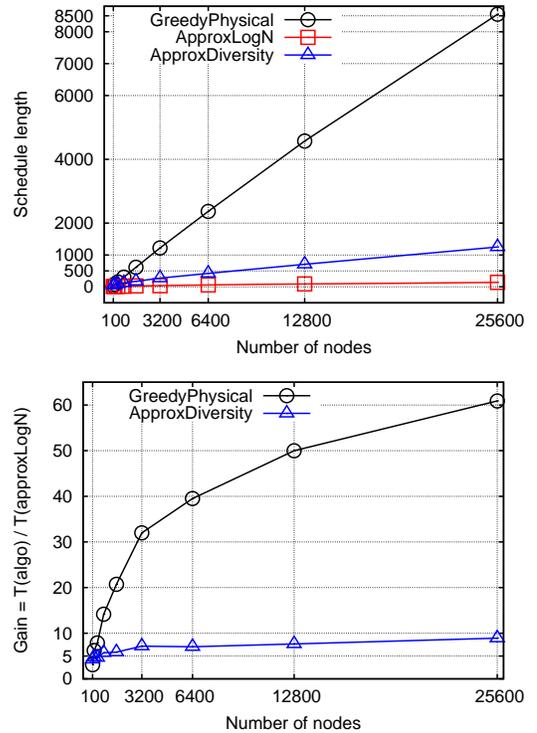

Fig. 3. Clustered Topology: $n_C = n/10$, $r_C = 10$.

In Figures 3(a) and 3(b) the results for the clustered topology are shown. As could be expected, the greedy algorithm is not able to deal with this more difficult scenario very efficiently. Even in very sparse topologies (100 nodes), GreedyPhysical computes 3 times longer schedules than ApproxLogN. As the density increases, the relative performance of the greedy algorithm deteriorates. ApproxLogN and ApproxDiversity compute even shorter schedules than in the random case, which indicates that they are able to schedule many clusters in parallel. The performance of ApproxLogN is still superior to that of ApproxDiversity.

In Figures 4(a) and 4(b) we analyze the influence of the cluster radius. In topologies with smaller clusters, i.e., in scenarios with higher density heterogeneity, the difference in performance becomes more accentuate. Whereas GreedyPhysical's performance slightly decreases with decreasing cluster radius, ApproxLogN and ApproxDiversity are able to compute ever shorter schedules. Smaller cluster radius means more separate clusters, which makes it easier to schedule clusters in parallel. GreedyPhysical, however, is not able to take advantage of this possibility. Among all three algorithms, ApproxLogN presents the best performance in all cases.

Next we analyze the influence of the path-loss exponent $\alpha$ in both random (Figures 5(a) and 5(b)) and clustered (Figures 6(a), and 6(b)) topologies. It can be seen that the performances of ApproxLogN and ApproxDiversity improve with increasing $\alpha$, whereas GreedyPhysical is more or less invariant to the path loss exponent. For $\alpha < 3$, in the random topology, GreedyPhysical presents a better performance than the other two algorithms. In the clustered topology, however, its performance is very poor even for low $\alpha$ and deteriorates relative to the other two approaches with increasing $\alpha$ in both kinds of topologies. Among all three algorithms, ApproxLogN presents the best performance for all values of $\alpha$ in the clustered topology and for $\alpha \geq 3$ in the random case.

To sum up, the simulations show that ApproxLogN, besides having an exponentially better analytical approximation ratio, presents advantages in challenging practical scenarios, such as high-density and heterogeneous-density networks.

## VII. Non-Approximability in Abstract SINR

In this section, we show that scheduling is extremely hard if the path-loss function can be non-geometric.

We distinguish "abstract SINR" ($SINR_A$) from "geometric SINR" ($SINR_G$) model according to the freedom with which the gain (or path-loss) matrix can be defined. In the $SINR_A$ model, as opposed to the $SINR_G$ model, path-loss between nodes is not constrained by their Euclidean coordinates, but can be set arbitrarily (i.e., triangular inequality need not be preserved when defining the path-loss matrix). Note that $SINR_A$ is more general and therefore a "harder" model than $SINR_G$, which we have been using to derive the results in the previous sections. We also remark that these results do not depend on complications due to noise.

*Theorem 7.1:* The scheduling problem in the $SINR_A$ model is at least as hard to approximate as the graph coloring problem, and the single-shot scheduling problem is as hard as the maximum independent set problem in graphs. In particular, the scheduling problem is NP-hard to approximate within $n^{1-\varepsilon}$-factor, for any $\varepsilon > 0$.

*Proof:* Let $G = (V, E)$ be a graph on $n$ vertices. We form an instance $I$ to the scheduling problem, containing a link $\ell_i$

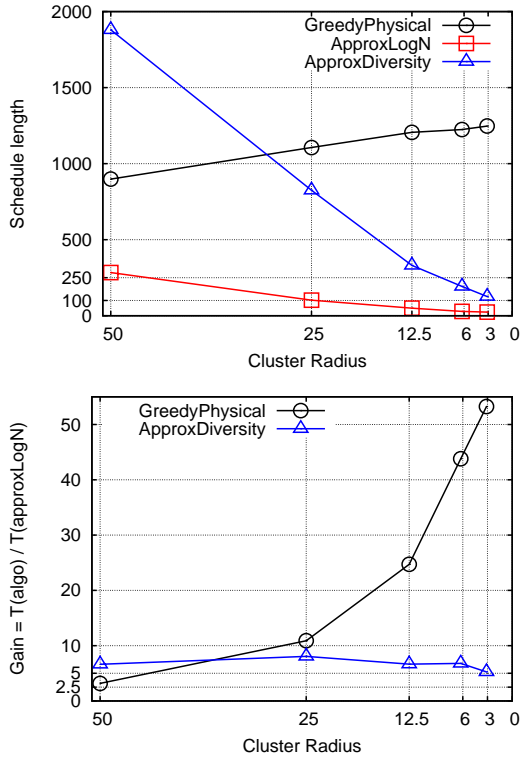

Fig. 4. Clustered Topology: $n = 3.2K$, $n_C = n/10$.

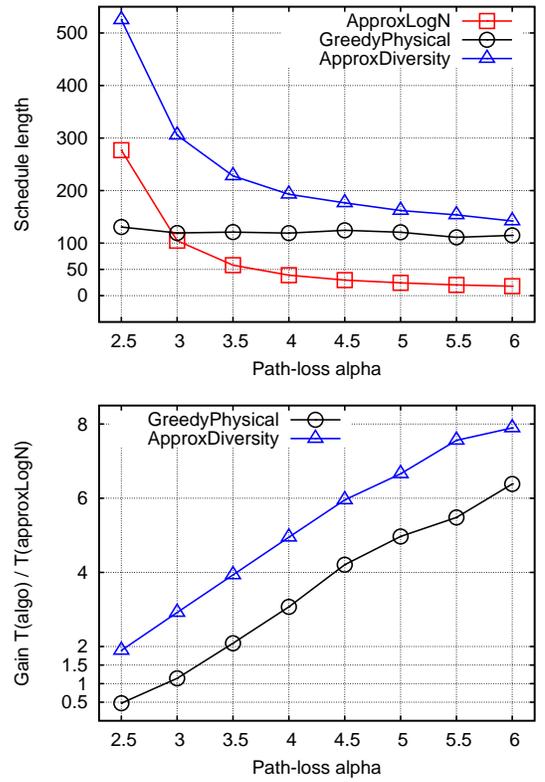

Fig. 5. Random topology: $n = 3.2K$, $l_{max} = 20$.

for each node $v_i$ and a symmetric gain matrix $A = (a_{ij})$. The value of $a_{ij}$ corresponds to the affectance of $\ell_i$ on $\ell_j$ (and, by symmetry, the affectance of $\ell_j$ on $\ell_i$). We define

$$a_{ij} = \begin{cases} 2 & \text{if } (v_i, v_j) \in E, \\ 1/n & \text{if } (v_i, v_j) \notin E. \end{cases}$$

Consider an independent set $S$ in $G$ and let $S'$ be the corresponding set of links in $I$. Observe that for any $\ell_v \in S'$, $a_{S'}(\ell_v) = (|S'| - 1) \cdot \frac{1}{n} < 1$, and thus $S'$ is feasible. Similarly, in any feasible set of links there can be no pair that correspond to adjacent vertices in $G$. It follows that there is one-to-one correspondence between independent sets in $G$ and feasible linksets in $I$. Hence, approximation algorithms for single-slot scheduling (scheduling) yield equivalent performance guarantees for the maximum independent set (minimum coloring) problem in graphs, respectively.

The last claim follows from the approximation hardness of graph coloring of [14], [40]. ∎

## VIII. CONCLUSIONS

The main open question is to obtain a constant factor approximation to the scheduling problem, as originally (and erroneously) claimed in [25]. Additionally, various parameter combinations are still open, and deserve more research, e.g. multi-hop traffic, scheduling and routing, analog network coding, stochastic fading models beyond pure geometric gain, such as Rician or Rayleigh fading.

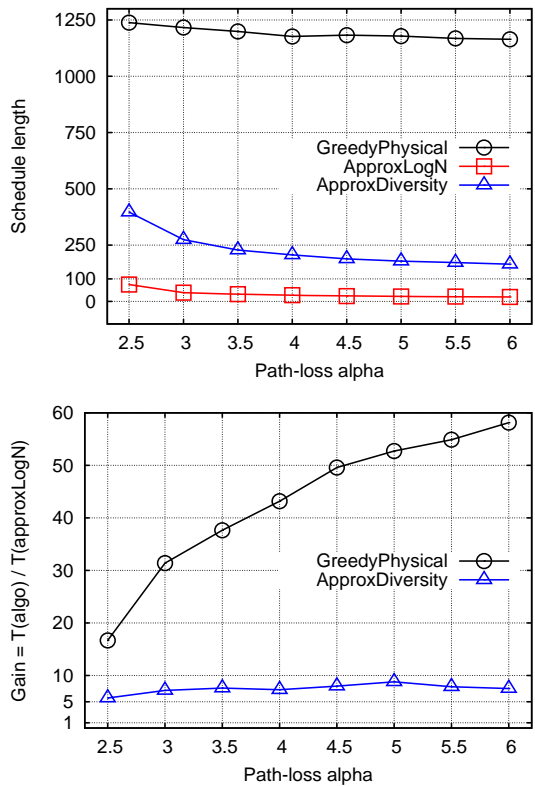

Fig. 6. Clustered topology: $n = 3.2K$, $n_C = n/10$, $r_C = 10$.